\documentclass[seceq,preprint]{ptptex}

\title{
Decay properties of the heavy-light mesons%
}

\author{
Takayuki \textsc{Matsuki},$^{1,}$
\footnote{E-mail: matsuki@tokyo-kasei.ac.jp;
an invited talk at  "New Frontiers in QCD 2010" held at Kyoto.}
and 
Koichi \textsc{Seo}$^{2,}$\footnote{E-mail: seo@gifu-cwc.ac.jp}
}

\inst{
$^1$Tokyo Kasei University, Tokyo 173-8602, Japan \\
$^2$Gifu City Women's College, Gifu 501-0192, Japan \\
}

\abst{
We study the decay properties of a heavy-light meson. We reformulate
the decay amplitudes for the heavy-light systems and find a new way to calculate
decay rates. Applying this formulation, we find a new sum rule
for the radiative decays of one heavy-light
meson into another, $H_1\to H_2+\gamma$ with various combinations of $H_i$.
}

\begin{document}

\maketitle

\section{Introduction}

Ever since the discovery of the so-called $D_{sJ}$ in 2003 by BaBar\cite{BB03} and
CLEO,\cite{CLEO03} which are now identified as $D_s(2317)$ and $D_s(2460)$ with $j_q^P
=(1/2)^+$ and $(3/2)^+$, respectively, the particles are discovered in the vicinity of
this energy or above one after another.
Theorists as well as experimentalists are trying to explain these particles as
ordinary $Q\bar q$, or molecular, or tetra-quark, or a bound state of diquark states,
{\it etc}.\cite{Review10}
At the present time there are several explanations for the same state and the final
decisive answer is not yet given.

As well as explanations of its origin, many people study the properties of
these heavy mesons, spin, parity, decay rates, etc.
In this paper, we report our study on the formulation to calculate decay rates of
a heavy-light meson which are Lorentz-invariant and on the kinematical
results derived from it.
We assume that the $D_{sJ}$ are heavy-light composite mesons $Q\bar q$, whose assumption is
supported by experimentally compatible numerical calculations of the lattice gauge
theory.\cite{PACS09,LIU10}

There are several ways to calculate decay rates of the heavy-light mesons proposed
so far; one is by Goity and Roberts\cite{GR99} which is applied to
newly discovered heavy mesons by Di Pierro and Eichten\cite{PE01}, \ and another by
Bardeen, Eichten, and Hill.\cite{BEH01}
The former is based on the ordinary dipole expansion of a bound state with the heavy
quark at rest at the center and the latter utilizes the effective chiral Lagrangian
which describes coupling of heavy-light mesons with chiral particles.

We propose a different method from these to calculate decay rates, which are
Lorentz-invariant and are calculated in the Breit frame where initial
and final mesons have the same velocity in the opposite directions.

\section{Formulation}

We start to construct formulation how to calculate a decay rate of a bound state
because the former methods are not Lorentz-covariant\cite{GR99,PE01} or do not
use meson wave functions to obtain the numerical results. We would like to solve these
problems in this paper.
For instance, consider the method proposed in Refs.~\citen{GR99} and \citen{PE01},
in which, {\it e.g.}, pion decay is described by the following equation.
\begin{eqnarray}
g{k_\mu }\int {{d^3}x\;\psi _f^\dag \left( {\vec x} \right){O^\mu }{\psi _i}} \left( {\vec x} \right){\exp\left({i\vec k \cdot \vec x}\right)},
\label{eq:piondecay}
\end{eqnarray}
where the interaction is assumed to be
\begin{eqnarray}
{{\cal L}_{{\mathop{\rm int}} }} = ig{\partial ^\mu }\phi {j_{5\mu }},
\end{eqnarray}
with axial vector current $j_{5\mu}$. Here one uses the composite meson wave functions,
$\psi_{i,f}$,
at the rest frame even for the final state $\psi_f$.
Another point is that a plane wave pion wave function
$\exp\left({i\vec k \cdot \vec x}\right)$ is inserted {\it ad hoc}.
Without phase factors in the initial and final wave functions
$\exp\left(-iP_{i,f}\cdot x\right)$,
the expression of Eq. (\ref{eq:piondecay}) neglects the {\it recoil} of heavy-light mesons.
We claim the factor in Eq. (\ref{eq:piondecay}) should be replaced with the
following expression in the Breit frame.
\begin{eqnarray}
  \exp \left( { - ik \cdot x} \right) \to \exp \left( { - 2i{m_Q} Vz} \right),
  \label{eq:translation}
\end{eqnarray}
where $m_Q$ is a heavy quark mass.
The factor $2$ on the r.h.s. of Eq. (\ref{eq:translation}) in the exponent appears
because the
initial heavy quark has a momentum $m_Q\gamma V$ in the $+z$ direction and the
final one has $m_Q\gamma V$ in the $-z$ in the Breit frame, hence the recoil momentum
becomes
$2m_Q\gamma V$. The factor $\gamma$ is absorbed into integral variable $z'=\gamma z$.

Both sides of Eq. (\ref{eq:translation}) are equal to each other
when one replaces $m_Q$ on the r.h.s. with an average of hadron masses as
$(m_1+m_2)/2$.
When we include  phase factors of initial and final wave functions $\psi_{i,f}$,
together with the plane wave pion wave function,
we obtain four-delta function meaning four-momentum conservation.
That is, the plane wave pion wave function disappears from the final expression but
there remains the recoil phase factor.

Our derivation of Eq. (\ref{eq:translation}) is based on the field theory and is
given as follows.
\begin{eqnarray}
&& \left\langle {0\left| {{q^c}\left( {t,\vec x} \right)Q\left( {t,\vec y} \right)} \right|P} \right\rangle  = \left\langle {0\left| {{q^c}\left( {0,\vec x - {{\vec X}_\xi }} \right)Q\left( {0,\vec y - {{\vec X}_\xi }} \right)} \right|P} \right\rangle {e^{ - iP \cdot {X_\xi }}}
\nonumber \\ 
&& \quad  = {\psi ^{(\xi )}}\left( {\vec x - \vec y;P} \right){e^{ - iP \cdot {X_\xi }}},\;{X_\xi } = \xi x + (1 - \xi )y,
\end{eqnarray}
where $\xi=0$ corresponds to $X_\xi$ being the heavy quark coordinate while $\xi=1$
the light quark one.

When calculating the transition amplitudes for $\xi=0$ and $1$, one obtains,
\begin{eqnarray}
&& \int {{d^4}x\;\left\langle {P',k\left| {{{\cal L}_{{\mathop{\rm int}} }}} \right|P} \right\rangle }
\nonumber  \\ 
&&  = {(2\pi )^4}{\delta ^4}\left( {P - P' - k} \right)\int {{d^3}z\;{\rm{tr}}\left[ {{\psi ^{'(0)\dag }}\left( {\vec z;P'} \right)O{\psi ^{(0)}}\left( {\vec z;P} \right)} \right]} \;{e^{ - i\vec k \cdot \vec z}} \label{xi:1} \\ 
&&  = {(2\pi )^4}{\delta ^4}\left( {P - P' - k} \right)\int {{d^3}z\;{\rm{tr}}\left[ {{\psi ^{'(1)\dag }}\left( {\vec z;P'} \right)O{\psi ^{(1)}}\left( {\vec z;P} \right)} \right]}.
\label{xi:2}
\end{eqnarray}
These expressions seem to be different and contradict from each other. However,
when one rewrites Eqs.~(\ref{xi:1}, \ref{xi:2}) in terms of the rest frame wave functions,
these become the same expression given by
\begin{eqnarray}
\left( G^{ - 1} OG \right)^{\alpha \beta } 
\int {d^3}z'\;\psi _{\alpha \gamma }^{'\dag } \left( \vec z~';M \right)
\psi_\beta {}^\gamma \left( \vec z~';M \right)\;e^{ - 2im_QVz'_3} ~.
\end{eqnarray}
where $G$ is a Lorentz transformation matrix.

The rest of our formulation is based on our former paper\cite{MS07} in which it is
shown how to
construct the Lorentz-boost wave function from the rest frame wave function.
In that paper, it is also shown that the Breit frame only with
$t'=x^{0'}=y^{0'}$, {\it i.e.}, times of the final heavy quark $Q(y^{0'},\vec y~')$
and light anti-quark $\bar q(x^{0'},\vec x~')$
are equal to each other, gives the Lorentz-invariant results, and hence
in this paper we also adopt this frame.
We study main decay modes of the heavy-light mesons, that is, one heavy-light
meson decays into another with one chiral particle or one photon (radiative decay),
{\it e.g.}, $D_{s}(0^+)\to D_s(0^-)+\pi^0$ or $D_{s}(0^+)\to D_s(1^-)+\gamma$.
The amplitudes for each process can be written in terms of form factors which
are given in the next section.

\section{Form Factors}

In this section, we study the form factors with an initial state of spin up to $2^+$.
In the case of radiative decays, amplitudes are given in terms of thirty electromagnetic
form factors as follows (if one takes into account the gauge invariance, the number of
independent form factors is reduced to twenty-three.) :
\begin{eqnarray}
  \frac{\left<0^-\left|j_\mu\right|1^-\right>}{i\sqrt{M_2M_1}} &=&
  \epsilon_{\mu\nu\rho\sigma}v_1^\nu v_2^\rho\epsilon^\sigma\xi_V^{(1)}, \label{AmpV1} \\
  \left<0^-\left|j_\mu\right|0^+\right> &=& 0, \\
  \frac{\left<1^-\left|j_\mu\right|0^+\right>}{i\sqrt{M_2M_1}} &=&
  (\omega+1)\epsilon_\mu^*\xi_{V1}^{(2)}+(\epsilon^*\cdot v_1)\left[
  \left(v_1+v_2\right)_\mu\xi_{V2}^{(2)}+\left(v_1-v_2\right)_\mu\xi_{V3}^{(2)}\right], \\
  \frac{\left<0^-\left|j_\mu\right|1^+\right>}{i\sqrt{M_2M_1}} &=&
  (\omega+1)\epsilon_\mu\xi_{V1}^{(i)}+(\epsilon\cdot v_2)\left[
  \left(v_1+v_2\right)_\mu\xi_{V2}^{(i)}+\left(v_1-v_2\right)_\mu\xi_{V3}^{(i)}\right],
  \nonumber \\
  \qquad &&~~(i=3,4) \\
  \frac{\left<1^-\left|j_\mu\right|1^+\right>}{\sqrt{M_2M_1}} &=&
  \epsilon_{\mu\nu\rho\sigma}\left[\epsilon_1^\nu\epsilon_2^{*\rho}\left\{
  \left(v_1+v_2\right)^\sigma\xi_{V1}^{(i)}+\left(v_1-v_2\right)^\sigma\xi_{V2}^{(i)}
  \right\} \right.
  \nonumber \\
  \qquad && \left. + \epsilon_1^\nu\left(\epsilon_2^*\cdot v_1\right)v_1^\rho
  v_2^\sigma\xi_{V3}^{(i)}+ \epsilon_2^{*\nu}\left(\epsilon_1\cdot v_2\right)v_1^\rho
  v_2^\sigma\xi_{V4}^{(i)} \right]
  \nonumber \\
  \qquad && + 
\epsilon_{\alpha\beta\gamma\delta}
\epsilon_1^\alpha\epsilon_2^{*\beta}v_1^\gamma v_2^\delta\left[
  \left(v_1+v_2\right)_\mu\xi_{V5}^{(i)}+\left(v_1-v_2\right)_\mu\xi_{V6}^{(i)}
  \right], ~~(i=5,6) \label{onepp} \\
  \frac{\left<0^-\left|j_\mu\right|2^+\right>}{\sqrt{M_2M_1}} &=&
  \epsilon_{\mu\nu\rho\sigma}\epsilon^{\nu\alpha}v_{2\alpha} 
  v_1^\rho v_2^\sigma\xi_V^{(7)}(\omega), \\
  \frac{\left<1^-\left|j_\mu\right|2^+\right>}{i\sqrt{M_2M_1}} &=&
  (\omega+1)\epsilon_{\mu\alpha}\epsilon^{*\alpha}\xi_{V1}^{(8)}
  +\epsilon_{\mu\alpha} v_2^\alpha(\epsilon^*\cdot v_1)\xi_{V2}^{(8)}
  +\epsilon_\mu^*\epsilon_{\alpha\beta} v_2^\alpha v_2^\beta\xi_{V3}^{(8)}
  \nonumber \\
  \qquad && + \epsilon_{\alpha\beta}\epsilon^{*\alpha} v_2^\beta\left[
  \left(v_1+v_2\right)_\mu\xi_{V4}^{(8)}+\left(v_1-v_2\right)_\mu\xi_{V5}^{(8)}
  \right]
  \nonumber \\
  \qquad && + \epsilon_{\alpha\beta}v_2^\alpha v_2^\beta(\epsilon^*\cdot v_1)\left[
  \left(v_1+v_2\right)_\mu\xi_{V6}^{(8)}+\left(v_1-v_2\right)_\mu\xi_{V7}^{(8)}
  \right], \label{AmpV8}
\end{eqnarray}
where $j_\mu=(-e_q)\overline{q}^c\gamma_\mu q^c+e_Q\overline{Q}\gamma_\mu Q$
with $e_q$ and $e_Q$ being electric charges of $q$ and $Q$, respectively.
In the case of decays with one chiral particle, amplitudes are given in terms of
twenty-six chiral particle form factors as follows:
\begin{eqnarray}
\frac{{\left\langle {{0^ - }\left| {{j_{5\mu }}} \right|{1^ - }} \right\rangle }}{{\sqrt {{M_2}{M_1}} }} &=&
(\omega+1)\, {\epsilon _{1\mu }}
\xi _{A1}^{(1)}  
\nonumber \\
\qquad &&+\left( {{\epsilon _1}\cdot{v_2}} \right)\left[ {{{\left( {{v_1} + {v_2}} \right)}_\mu }\xi _{A2}^{(1)} + {{\left( {{v_1} - {v_2}} \right)}_\mu }\xi _{A3}^{(1)}} \right], \\
\frac{{\left\langle {{0^ - }\left| {{j_{5\mu }}} \right|{0^ + }} \right\rangle }}{{i\sqrt {{M_2}{M_1}} }} &=& {\left( {{v_1} + {v_2}} \right)_\mu }\xi _{A1}^{(2)} + {\left( {{v_1} - {v_2}} \right)_\mu }\xi _{A2}^{(2)}, \\
\frac{{\left\langle {{1^ - }\left| {{j_{5\mu }}} \right|{0^ + }} \right\rangle }}{{\sqrt {{M_2}{M_1}} }} &=& {\epsilon_{\mu \nu \rho \sigma }}v_1^\nu v_2^\rho \epsilon_2^{*\sigma }{\xi _{A}^{(3)}}, \\
\frac{{\left\langle {{0^ - }\left| {{j_{5\mu }}} \right|{1^ + }} \right\rangle }}{{\sqrt {{M_2}{M_1}} }} &=& {\epsilon _{\mu \nu \rho \sigma }}v_1^\nu v_2^\rho \epsilon _1^\sigma {\xi _{A}^{(i)}},\quad (i = 4,5) \\
 \frac{{\left\langle {{1^ - }\left| {{j_{5\mu }}} \right|{1^ + }} \right\rangle }}{{i\sqrt {{M_2}{M_1}} }} &=& \left( {\epsilon _2^* \cdot {\epsilon _1}} \right){\left( {{v_1} + {v_2}} \right)_\mu }\xi _{A1}^{(i)} + \left( {\epsilon _2^* \cdot {\epsilon _1}} \right){\left( {{v_1} - {v_2}} \right)_\mu }\xi _{A2}^{(i)} \nonumber \\ 
\qquad &&  + \left( {\epsilon _2^* \cdot {v_1}} \right){\epsilon _{1\mu }}\xi _{A3}^{(i)} + \left( {{\epsilon _1} \cdot {v_2}} \right)\epsilon _{2\mu }^*\xi _{A4}^{(i)}, \quad (i=6,7) \\
\frac{{\left\langle {{0^ - }\left| {{j_{5\mu }}} \right|{2^ + }} \right\rangle }}{{i\sqrt {{M_2}{M_1}} }} &=& {\epsilon _{1\mu \nu }}v_2^\nu \xi _{A1}^{(8)} + {\epsilon _{1\alpha \beta }}v_2^\alpha v_2^\beta \left( {{v_{1\mu }}\xi _{A2}^{(8)} + {v_{2\mu }}\xi _{A3}^{(8)}} \right), \\
%
 \frac{{\left\langle {{1^ - }\left| {{j_{5\mu }}} \right|{2^ + }} \right\rangle }}{{\sqrt {{M_2}{M_1}} }} &=& {\epsilon_{\mu \nu \rho \sigma }}\left[ {\epsilon_1^{\nu \alpha }\epsilon_2^{*\rho }{v_{2\alpha }}\left\{ {{{\left( {{v_1} + {v_2}} \right)}^\sigma }\xi _{A1}^{(9)} + {{\left( {{v_1} - {v_2}} \right)}^\sigma }\xi _{A2}^{(9)}} \right\}} \right.
\nonumber \\
\quad && \left. { + v_1^\rho v_2^\sigma \left\{ {\epsilon _1^{\nu \alpha }\epsilon _{2\alpha }^*\xi _{A3}^{(9)} + \epsilon _1^{\nu \alpha }{v_{2\alpha }}\left( {\epsilon _2^*\cdot{v_1}} \right)\xi _{A4}^{(9)} + {\epsilon _{1\alpha \beta }}v_2^\alpha v_2^\beta \epsilon _2^{*\nu }\xi _{A5}^{(9)}} \right\}} \right]
\nonumber \\ 
\quad &&  + {\epsilon_{\alpha \beta \gamma \delta }}
\epsilon_1^{\alpha \lambda }
{v_{2\lambda }}
\epsilon_2^{*\beta }
v_1^\gamma v_2^\delta \left[ {{{\left( {{v_1} + {v_2}} \right)}_\mu }\xi _{A6}^{(9)} + {{\left( {{v_1} - {v_2}} \right)}_\mu }\xi _{A7}^{(9)}} \right].
\label{AmpA8}
\end{eqnarray}
Here $\epsilon_{\mu\nu}$ and $\epsilon_\mu$ are polarization vectors, which could have
subindex to distinguish whether it is for initial $(i=1)$ or final $(i=2)$.
All form factors depend only on $\omega=(v_1\cdot v_2)$, where $v_i$ are initial and
final velocities of heavy-light mesons, $H_i$ $(H_1 \to H_2+\pi/\gamma)$.
These amplitudes can be written in terms of form factors {\it a la} semi-leptonic
decay of a heavy-light meson. These are, however, not described in terms of just one
form factor like the Isgur-Wise function because the interaction can not be regarded
as a point contrary to the semi-leptonic decay.

All amplitudes given above Eqs.~(\ref{AmpV1})-(\ref{AmpA8}) are in principle calculable
at the lowest order in $1/m_Q$ because we know the explicit forms of wave functions of
any spin in the heavy quark symmetric limit\cite{Matsuki97,Matsuki05} which
are described in the next section.

\section{Sum Rule}

In this section, we show one of our results for the radiative decay of a heavy-light
meson. By analytically calculating amplitudes given by
Eqs.~(\ref{AmpV1})-(\ref{AmpV8}), we have the following sum rule in the heavy
quark symmetric limit:
\begin{eqnarray}
&& \Gamma \left( {{\,``^3}P{}_1"\left(1^+\right) \to {\,^1}S{}_0\left(0^-\right)  + \gamma } \right) + \Gamma \left( {{\,``^3}P{}_1"\left(1^+\right)  \to {\,^3}S{}_1\left(1^-\right) + \gamma } \right) 
\nonumber  \\ 
&& \quad =
 \Gamma \left( {{\,^3}P{}_0\left(0^+\right)  \to {\,^3}S{}_1\left(1^-\right) + \gamma } \right) 
= \frac{2}{9}\frac{{{e^2}{k^3}}}{{2\pi }}{\left[ {\int {dr\left( {u_2^{ - 1}u_1^1 + %
v_2^{ - 1}v_1^1} \right)} } \right]^2} \label{sum1} \\ 
&& \Gamma \left( {{\,``^1}P{}_1"\left(1^+\right)  \to {\,^1}S{}_0\left(0^-\right)  + \gamma } \right) + \Gamma \left( {{\,``^1}P{}_1"\left(1^+\right)  \to {\,^3}S{}_1\left(1^-\right) + \gamma } \right)
\nonumber  \\ 
&& \quad =
 \Gamma \left( {{\,^3}P{}_2\left(2^+\right) \to {\,^3}S{}_1\left(1^-\right) + \gamma } \right) 
 = \frac{2}{9}\frac{{{e^2}{k^3}}}{{2\pi }}{\left[ {\int {dr\left( {u_2^{ - 1}u_1^{ - 2} + v_2^{ - 1}v_1^{ - 2}} \right)} } \right]^2}. \label{sum2}
\end{eqnarray}
In the heavy quark symmetric Hamiltonian, the light quark
degrees of freedom are conserved and the $LS_q$ couling is included.
Hence, the mass eigenstates in the heavy quark symmetric limit
become mixed states of pure states\cite{Matsuki97},\cite{Matsuki10-1} and are given by
\begin{eqnarray}
\left( {\begin{array}{*{20}{c}}
   {\left| ``{^3{P_1}}" \right\rangle }  \\
   {\left| ``{^1{P_1}}" \right\rangle }  \\

 \end{array} } \right) = \frac{1}{\sqrt{3}} \left( {\begin{array}{*{20}{c}}
   {\sqrt{2} } & { -1 }  \\
   { 1 } & {\sqrt{2} }  \\

 \end{array} } \right)\left( {\begin{array}{*{20}{c}}
   {\left| {^3{P_1}} \right\rangle }  \\
   {\left| {^1{P_1}} \right\rangle }  \\

 \end{array} } \right).
\end{eqnarray}
In Eqs. (\ref{sum1}, \ref{sum2}),
$u_i^k(r)$ and $v_i^k(r)$ are initial $(i=1)$ and final $(i=2)$ radial upper/lower
parts of the wave functions with a quantum number $k$ as explained below.
The wave function, which has two spinor indices and is expressed as $4\times 2$
matrix form, is explicitly given by
\begin{eqnarray}
  \Psi _{j\,m}^k(\vec r) &=& \sqrt{\frac{2M}{4\pi}}\frac{1}{r}\left( 
\begin{array}{c}
u^k(r) \\ 
-iv^k(r)\left(\vec\sigma\cdot\vec n\right)
\end{array}
\right) \;y_{j\,m}^k,
  \label{0thsols2}
\end{eqnarray}
where $\vec n=\vec r/r$, $y_{j\,m}^k$ being of $2\times 2$ form are the angular part
of the wave
function, and $u^k(r)$ and $v^k(r)$ are the radial parts. The total angular
momentum of a heavy meson $\vec J$ is the sum of the total angular momentum of
the light quark degrees of freedom $\vec j_q$ and the heavy quark spin
${1\over 2}\vec \Sigma_Q$:
\begin{equation}
  {\vec J} = {\vec j_q} +{1\over 2}\vec \Sigma_Q\qquad {\rm with} \quad
  {\vec j_q}=\vec L + {1\over 2}\vec \Sigma_{\bar q},
\end{equation}
where ${1\over 2}\vec\Sigma_{\bar q}\ (={1\over 2}
\vec\sigma_{\bar q}\;1_{2\times 2})$ and $\vec L$ are the $4\times 4$ spin and
the orbital angular momentum of a light antiquark, respectively. Furthermore,
$k$ is the quantum number of the following spinor operator $K$ \cite{Matsuki97,MMMS04}
\begin{equation}
  K = -\beta_{\bar q} \left( \vec \Sigma_{\bar q} \cdot \vec L + 1 \right),
  \qquad
  K\, \Psi_{j\,m}^k = k\, \Psi_{j\,m}^k.
  \label{k_quantum}
\end{equation}
$j_q$ and $k$ are good quantum numbers in the heavy quark limit, hence the
wave function can be described in these two.

Applying Eqs.~(\ref{sum1}) and (\ref{sum2}) to actual heavy-light mesons, e.g., 
to $D_s$, we have
\begin{eqnarray}
&& \Gamma \left( {{\mkern 1mu} {D_{s1}}(2460) \to {D_s}(1968) + \gamma } \right) + \Gamma \left( {{\mkern 1mu} {D_{s1}}(2460) \to {\mkern 1mu} D_{s1}^*(2112) + \gamma } \right) 
\nonumber \\
&& \qquad =\Gamma \left( {{D_{s0}}(2317) \to D_{s1}^*(2112) + \gamma } \right), \\
&& \Gamma \left( {{D_{s1}}(2536) \to {D_s}(1968) + \gamma } \right) + \Gamma \left( {{D_{s1}}(2536) \to D_{s1}^*(2112) + \gamma } \right) 
\nonumber \\ 
&& \qquad  = \Gamma \left( {{D_{s2}}(2573) \to D_{s1}^*(2112) + \gamma } \right),
\end{eqnarray}
where experiments give
\begin{eqnarray}
&& \Gamma \left({{\mkern 1mu}{D_{s1}}(2460)\to{D_s}(1968)+\gamma}\right)< 630 
~{\rm keV}, \\
&& \Gamma \left( {{\mkern 1mu} {D_{s1}}(2460) \to {\mkern 1mu} D_{s1}^*(2112) + \gamma } \right)<280 ~{\rm keV}, \\
&& \Gamma \left( {{D_{s0}}(2317) \to D_{s1}^*(2112) + \gamma } \right)
 ~~{\rm not~yet~seen}, \\
\nonumber \\
&& \Gamma \left({{\mkern 1mu}{D_{s1}}(2536)\to{D_s}(1968)+\gamma}\right)
~~{\rm not~yet~seen}, \\
&& \Gamma \left( {{\mkern 1mu} {D_{s1}}(2536) \to {\mkern 1mu} D_{s1}^*(2112) + \gamma } \right)~~{\rm possibly~seen}, \\
&& \Gamma \left( {{D_{s0}}(2573) \to D_{s1}^*(2112) + \gamma } \right)
 ~~{\rm not~yet~seen},
\end{eqnarray}
We have to wait and see to what extent the sum rules Eqs.~(\ref{sum1}, \ref{sum2})
are satisfied until experiments observe the above and other radiative decay modes.

\section{Summary}

We have proposed a new method given by Eq.~(\ref{eq:translation}) how to calculate
decay rates for a heavy-light
mesons and have applied it to the processes $H_1\to H_2+\phi^a/\gamma$ with
one chiral particle $\phi^a$ or one photon in the final state.
We have derived form factors of vector as well as axial vector currents, which can
be used to calculate the above processes.

We have derived the sum rules given by Eqs.~(\ref{sum1}) and (\ref{sum2}) for the
radiative decay rates in the heavy quark symmetric limit,
\begin{eqnarray}
 \sum\limits_{S = {0^ - },{1^ - }} {\Gamma \left( {``{^3}{P_1}"\left( {{1^ + }} \right) \to S + \gamma } \right)}  = \Gamma \left( {{0^ + } \to {1^ - } + \gamma } \right), \\ 
 \sum\limits_{S = {0^ - },{1^ - }} {\Gamma \left( {``{^1}{P_1}"\left( {{1^ + }} \right) \to S + \gamma } \right)}  = \Gamma \left( {{2^ + } \to {1^ - } + \gamma } \right),
\end{eqnarray}
which needs to be checked by future experiments.
We still need to give numerical results for all the processes in concern,
$H_1\to H_2+\phi^a/\gamma$, which will be published in near future.\cite{Matsuki10}


%


\begin{thebibliography}{99}
%
\bibitem{BB03} BaBar Collaboration, B. Aubert {\em et al.}, \PRL{90,2003,242001}.
\bibitem{CLEO03} CLEO Collaboration, D. Besson {\em et al.}, \PRD{68,2003,032002}.
\bibitem{Review10} See for review: E. S. Swanson, \PR{429,2006,243};
  L.-L. Shen {\it et al.}, arXiv:1005.0994.
\bibitem{PACS09} PACS-CS collaboration, Y. Namekawa {\em et al.}, arXiv:0810.2364, a talk given at the kick-off meeting of ``Elucidation of new hadrons with a variety of flavors"
held at Nagoya Univ., Nov. 27-28, 2009.
\bibitem{LIU10} K.F. Liu, a talk given on Feb. 5, 2010 at ``New Frontiers of in QCD 2010"
held at Yukawa Inst. for Theor. Phys. of Kyoto Univ., Jan. 18 through Mar. 19, 2010.
\bibitem{GR99} J.L. Goity and W. Roberts, \PRD{60,1999,034001}.
\bibitem{PE01} M. Di Perro and E.J. Eichten, \PRD{64,2001,114004}.
\bibitem{BEH01} W.A. Bardeen, E.J. Eichten, and C.T. Hill \PRD{68,2003,054024}.
\bibitem{MS07} T. Matsuki and K. Seo, \PTP{118,2007,1}.
\bibitem{Matsuki97} T. Matsuki and T. Morii, \PRD{56,1997,5646}
\bibitem{Matsuki05} T. Matsuki, T. Morii, and K. Sudoh, \PTP{117,2007,1077}.
\bibitem{Matsuki10-1} T. Matsuki, T. Morii and K. Seo,
Prog. Theor. Phys. {\bf 124} (2010), 285, arXiv:1001.4248.
\bibitem{MMMS04} T. Matsuki, T. Mawatari, T. Morii, K. Sudoh, hep-ph/0408326.
\bibitem{Matsuki10} T. Matsuki and K. Seo, work in progress.
\end{thebibliography}
\end{document}